\documentstyle[aps]{revtex}
\def\be{\begin{equation}}
\def\ee{\end{equation}}
\begin{document}
\title{Vortex Plasma in a Superconducting Film with Magnetic Dots}
\author{D.E. Feldman$^{1,2}$, I.F. Lyuksyutov$^1$ 
and V.L. Pokrovsky$^{1,2}$}
\address{$^1$Department of Physics, Texas A\&M University, College Station,
Texas 77843-4242 \\
$^2$Landau Institute for Theoretical Physics, Chernogolovka,
Moscow region 142432, Russia}
\maketitle

\begin{abstract}
We consider a superconducting film, placed upon a magnetic dot array.
Magnetic moments of the dots are normal to the film and randomly oriented.
We determine how the concentration of the vortices in the film  depends
on the magnetic moment of a dot at low temperatures. The concentration
of the vortices, bound to the dots, is proportional to the density
of the dots and depends on the magnetization of a dot in a step-like way.
The concentration of the unbound vortices oscillates about a value,
proportional to the magnetic moment 
of the dots. The period of the oscillations
is equal to the width of a  step in the concentration
of the bound vortices.
\end{abstract}



Superconductivity of thin films was studied for a long
time \cite{1}. An important difference  of the two-dimensional
superconductors from the three-dimensional ones is related
with the topological defects. Vortices appear
in thin films of the superconductors  which are of the first
kind in the bulk \cite{2}.
They can appear spontaneously even in the absence of the
magnetic field. In specially prepared films with the size about
the effective screening length, unbound vortices appear above the
Berezinskii-Kosterlitz-Thouless transition \cite{3}-\cite{3a}.

A recent surge of interest to this problem is
associated with advances in preparation of magnetic nanostructures
interacting with the superconducting films 
\cite{4a}-\cite{4c}. 
Magnetic field from the magnetic nanostructures (dots)
gives rise to vortices and pins them.  
As a result, in a superconducting film placed upon an
array of magnetic dots, a periodic field dependence of the 
magnetoresistance and  superconducting transition temperature
was observed \cite{4}, \cite{4b}.

Theoretically the problem of a superconducting film 
supplied by a periodic array of magnetic 
dots was studied in Ref. \cite{5}. It was shown that the 
properties of such a system depend on the orientation of the 
magnetization of the 
dots, their mutual distances and the coercive field. In the case of
strong coercive force, the reorientation of the magnetic moments
is a slow process. 
Hence,  a random array of magnetic moments occurs at zero-field cooling
below the Curie temperature.
 If the easy magnetic axes
are perpendicular to the film,
each dot favors creation of a vortex \cite{5}. 
Thus, a random
vortex structure appears in the superconductor.
The vortices pinned by the dots induce a random potential 
in the film. If the period of the dot array is small enough,
the random potential may be sufficient for creation of 
additional unpinned vortices.  
The resulting vortex plasma phase is expected to have a much 
larger resistance, than the pinned vortex state \cite{5}.


In the present paper we consider a toy model of 
the vortex plasma state which reproduces the phenomena predicted
in \cite{5}, but leads also to new predictions
 of a rich phase diagram and elementary excitations.
We study the dependence of the concentration of the vortices on the
magnetic moment of a dot. The concentration of the pinned vortices is 
found to be proportional to the density of the dots. The concentration
of the unbound vortices is less to the factor of order $a/\lambda_{eff}$,
where $\lambda_{eff}$ is the effective screening length \cite{1},
$a$ the lattice constant of the dots.
Both concentrations grow as the magnetic moment of a dot increases.
The concentration of the pinned vortices depends on the magnetic moment
in a step-like way. The concentration of the unpinned vortices 
has linear and oscillating components.

We consider first a thin film of the size $\sim\lambda_{eff}=\lambda^2/d$,
where $\lambda$ is the London penetration length, $d$ the thickness
of the film. Then we discuss what happens in a film, which
is larger than the effective screening length $\lambda_{eff}$.
We argue that the behavior of the vortex plasma does not 
depend qualitatively on the size of the system.

The energy of the system of the vortices is composed of
three components.
These are the single vortex energies, the energy of the 
vortex-vortex interaction and the coupling energy between the
vortices and the dots. A single vortex energy is \cite{1}

\be
\label{1}
 E_i=\epsilon_0 n_i^2\ln\frac{\lambda_{eff}}{\xi},
\ee
$n_i$ being the vorticity,  
$\xi$ the core size and
$\epsilon_0=\Phi_0^2/16\pi^2\lambda_{eff}$,
where $\Phi_0$ is the magnetic flux quantum.
The interaction of a pair of vortices
separated by a distance $r$ 
decreases fast at $r>\lambda_{eff}$,
while at $r<\lambda_{eff}$ it depends on the distance 
logarithmically \cite{6a}

\be
\label{2a}
E_{ij}=2\epsilon_0 n_i n_j
\ln\frac{\lambda_{eff}}{r},
\ee
where $n_i, n_j$ are the vorticities.  
The interaction of a vortex and a dot is estimated in Ref. 
\cite{5} as

\be
\label{3}
E_i^d=\epsilon_0\frac{\Phi_d}{\Phi_0}n_i,
\ee
where $\Phi_d$ is the magnetic flux generated by the dot.
The total Hamiltonian of the system reads

\be
\label{4}
H=\sum_i E_i +\sum_{i>j} E_{ij} + \sum_i E_i^d.
\ee

The vortex plasma appears when
the distance between the dots $a\ll\lambda_{eff}$ \cite{5}.
Then the number of the dots in the film of the size 
$\lambda_{eff}$ is $N_D=(\lambda_{eff}/a)^2\gg 1$. 
In this case the system, governed by the Hamiltonian (\ref{4}),
displays strong collective effects. 
For simplification, we approximate the slow logarithmic
dependence (\ref{2a}) at the scales $a<r<\lambda_{eff}$
with a constant. Then the model Hamiltonian
mimicking the real Hamiltonian (\ref{4})  
has a form

\be
\label{5}
H=-U\sum_i \sigma_i n_i +  \sum_i E_i + 2\epsilon_0\sum_{i> j}n_in_j,
\ee
where $\sigma_i=\pm 1$ describe two possible orientations 
of the magnetizations
of the dots, $U=\epsilon_0\Phi_d/\Phi_0$. 
This Hamiltonian can be rewritten as

\be
\label{6}
H=-U\sum_i \sigma_i n_i +\epsilon \sum n_i^2 + \epsilon_0
\left(\sum_{i} n_i\right)^2,
\ee
where $\epsilon=\epsilon_0\ln(a/\xi)$.

The minimum of the energy (\ref{6}) corresponds to zero
total vorticity $Q=\sum_i n_i$. Indeed, if $\Delta Q$
vortices are removed, so that the vorticity become $Q'=Q-\Delta Q$,
the last term in (\ref{6}) would decrease by the value proportional
to $Q\Delta Q$. At the same time, the maximal possible energy loss
due to the first two terms of the Hamiltonian is proportional to
$\Delta Q$. Hence, decreasing the total vorticity
is favorable unless $Q\sim 1$. In the system with the number of vortices
$\sim (\lambda_{eff}/a)^2\gg 1$ this is practically zero. Thus,
the ``neutrality'' condition must be satisfied:

\be
\label{cond}
\sum_i n_i=0. 
\ee

Our aim is to find the ground state of the Hamiltonian (\ref{6})
with the constraint (\ref{cond}). The ground state depends on the parameter
$\kappa=U/\epsilon$. The discussion below is limited by the case
$\kappa\ll \lambda_{eff}/a$.

Let us assume that there are $N$ ``positive'' dots  favoring creating 
positive vortices, and $N+K$ ``negative'' dots which favor
negative vortices. Obviously, $N\approx \lambda_{eff}^2/(2a^2)$ and
$K\sim \lambda_{eff}/a$ are random.
Note that in the
ground state all the unbound vortices have the same sign and the unit 
vorticity. Below we assume that the unbound vortices are positive.
As seen below, this assumption is equivalent to the condition
$K>0$.
The case of the negative vorticity is completely analogous. 


Let us consider an arbitrary dot with occupancy $n$. The neutrality
condition (\ref{cond}) allows to change the occupancy by $\pm 1$ and
simultaneously create an unbound vortex or antivortex. In the ground
state these excitations can not decrease the energy. This gives a
restriction on the possible values of the occupancy.
The energies of the excitations are 

\be
\label{bad1}
\Delta E = \epsilon[1+(n\pm 1)^2-(n\pm 1)\kappa] -\epsilon[n^2-n\kappa]=
\epsilon[2\pm (2n-\kappa)]\ge 0.
\ee
Hence, $2n-2\le\kappa\le 2n+2$.
Thus, at $\kappa=2(q+\delta)$, where $q$ is integer, $0<\delta<1$, 
the only possible values of $n$ are $q$ and $q+1$. At $\kappa=2q$
an additional possible occupancy is $q-1$.

Let us show that a non-zero number of unbound vortices appears when $\kappa$
is equal to an even integer only.
Indeed, for the system in the ground state, 
the energy must not decrease when an unbound vortex is placed onto a dot, 
or when an unbound vortex and a bound 
antivortex are removed.
%
%
The consideration, similar to the derivation of Eq. (\ref{bad1}), 
leads to a condition

\be
\label{8}
2m_{min}\ge\kappa\ge 2n_{max},
\ee
where $m_{min}$ and $n_{max}$
are the minimal occupancy of the positive dots
and the maximal occupancy of the negative ones
respectively.
Eq. (\ref{8}) is compatible with (\ref{cond})
only if the inequalities in (\ref{8}) are actually
equalities.
Indeed, the number of the positive vortices $V^+\ge Nm_{min}$
and the negative vortex number $V^-\le (N+K)n_{max}$.
Since $K\ll N$, it follows from the neutrality condition
$V^+=V^-$ that $m_{min}= n_{max}$.
Hence,  $\kappa=2k$ where $k$ is an integer.
Note that Eq. (\ref{8}) does not contradict Eq. (\ref{cond}) if $K\ge 0$
only. We assume below that this is the case.


Let us  consider the case of $\kappa=2(q+\delta)$, $0<\delta<1$.
At these values of $\kappa$ unbound vortices are absent. 
Let the number of the positive
dots with occupancy $q$ be $S$. The numbers of the negative dots
with occupancies $q$ and $q+1$ are  then determined by the neutrality 
condition. The energy as a function of $S$ is
$E(S)={\rm constant} + 2S\epsilon[\kappa-2q-1]$.
Depending on $\kappa$, the minimum of the energy $E(S)$
corresponds to $S=0$ or $S=N$. At $\kappa=2q+1$ the ground state is
degenerate.

Now we study the case of $\kappa=2q$.
We denote the numbers of the positive dots with occupancies
$q-1$ and $q+1$ as $A_+$ and $B_+$ respectively,
the numbers of the negative dots with occupancies
$q-1$ and $q+1$ as $A_-$ and $B_-$ respectively.
The number of the unbound vortices is determined by the
neutrality. The energy dependence on $A_\pm, B_\pm$
is given by the expression
$E={\rm constant}+2\epsilon[A_+ +B_-]$.
Thus, in the ground state $A_+ =B_- =0$.
At the same time the energy does not depend on $A_-$ and $B_+$.


Now we are in position to describe all the ground states.
Below we consider the case when the number of the negative
dots $(N+K)$ is larger than the number of the positive dots $N$.
The opposite case is analogous.

1) At $\kappa<1$ the vortices are absent.

2) At $\kappa=2n-1$ the ground state is degenerate.
All the dots can be divided into 4 groups with occupancies
$n$ positive, $(n-1)$ positive, $(n-1)$ negative and
$n$ negative vortices on each dot, respectively.
The numbers of dots in these groups are $N-S$, $S$,
$nK+S$ and $N-(n-1)K-S$, respectively, where $S$
is any integer satisfying inequality $S\le N-(n-1)K$.


3) At $2n-1<\kappa<2n$, $n$ vortices are bound with each positive dot
and with $N-(n-1)K$ negative dots. Each of the other $nK$ negative 
dots is occupied by $(n-1)$ vortices.

4) At $\kappa=2n$ the ground state is degenerate.
There are 4 groups of dots with  occupancies $(n-1)$ negative,
$n$ negative, 
$(n+1)$ positive and $n$ positive vortices
on each dot. The groups contain $S$, $N+K-S$,
$nK-S-P$ and $N+S+P-nK$ dots, respectively, where integer $P$
and $S$ satisfy inequality $S+P\le nK$. Besides, there are
$P$ unbound vortices.


5) At $2n<\kappa<(2n+1)$ each negative dot is occupied by
$n$ vortices. $nK$ positive dots are occupied by $(n+1)$
vortices and the other
$N-nK$ positive dots are occupied by $n$ vortices.

Thus, the concentration $c_b$ of the bound vortices obeys 
a step-like low

\be
\label{B}
c_b=\left[\frac{\kappa+1}{2} \right]
\frac{1}{a^2} + O\left(\frac{1}{\lambda_{eff}a}\right),
\ee
where the square brackets denote the integer part.
The unbound vortices exist only at $\kappa=2n$, and their concentration
$c_u$ satisfies a  relation

\be
\label{U}
c_u\sim\frac{n|K| }{\lambda_{eff}^2}
\ee
where $|K|$ is the
absolute value 
of the difference between the numbers of the positive and negative dots.
The disorder average of $|K|$ is $2\lambda_{eff}/(\sqrt{\pi}a)$.

The fact, that the concentration of the unbound vortices is proportional
to $1/(\lambda_{eff}a)$, can be understood in 
the framework of the approach
\cite{5}. It was argued in Ref. \cite{5} that the bound vortices
induce a random potential with the characteristic variation 
 $\epsilon_0\lambda_{ eff}/a$.
This leads to creation of unbound vortices screening the
random potential. The concentration at which appearance of
new vortices becomes unfavorable is $c_u\sim 1/(\lambda_{eff}a)$.

The model (\ref{5}) is oversimplified in two respects.
First, within the model the potential created by the vortices is
completely screened. It is natural that in the ground state
the potential is screened at the scales $r>a$, larger
than the intervortex distance. However, the potential can not 
be screened at the scales $r\le a$. This unscreened potential 
provides an additional contribution
to the Hamiltonian (\ref{5}) leading to dependence of 
the vortex energy 
on the position. This contribution lifts the degeneracy 
of the ground state at even $\kappa$ and fixes the number of the unbound
vortices. It also makes the creation of the unbound vortices favorable
at non-integer values of $\kappa$.  This is
a consequence of the fact
that the low-lying states are almost degenerate
at $\kappa\approx 2n$, and the distances between the low-lying 
levels 
may turn out to be less than the value of the unscreened 
potential variation. Still the maxima of the  unbound vortex density
correspond to the integer even values of $\kappa$.
Another effect of the incomplete screening is the lifting
of the ground state degeneracy at $\kappa=2n+1$.
In this case the total number of the bound vortices
is determined by the unscreened potential. This potential
smears the concentration steps at the odd integer values of   
$\kappa$.


The second simplification consists in the choice of the energy of
a vortex upon a dot in the form $E=E_i+E_i^d$, where $E_i$
and $E_i^d$ are given by Eqs. (\ref{1},\ref{3}). 
This value of $E$ provides only
an upper boundary for the energy of a vortex pinned by a dot.
Since the energies
of the bound vortices are lower than it is assumed in Eq. (\ref{5}),
the creation of the unbound vortices is less favorable.
As a consequence their concentration in a more realistic model is 
lower than 
(\ref{U}).

We expect that the behavior of a large film is qualitatively the same
as the behavior of the film of the size $\lambda_{eff}$. 
The interaction of the vortices at the distance $r>\lambda_{eff}$
can be calculated with the method of Ref. \cite{2}. The main
contribution originates from the interaction of the magnetic
fields, induced by the vortices. It depends on the distance
as $V\sim 1/r$. Due to the screening, the blocks of the size 
$\lambda_{eff}$ are to be considered not as free charges, but as
dipoles. Their potential obeys $1/r^2$ law. Since the orientation of the
dipoles is random, the interaction of the distant blocks
is irrelevant. Thus, to cut-off the intervortex interaction at some scale
$R\sim\lambda_{eff}$ is a reasonable approximation. 
Then a qualitatively
correct picture is given by the following model.
The system is divided into blocks of the size $R$. The
interaction of the vortices from the different blocks is neglected.
Inside a block the Hamiltonian (\ref{4}) is valid. The main features of this
model are the same as in the film of the size $\lambda_{eff}$. 
  
Besides the ground states, we have determined the spectrum
of elementary excitations. In particular, the energy cost
of an unbound vortex is $\epsilon_0|\kappa-2[(\kappa+1)/2]|$,
where $[...]$ denotes the integer part. 
Our approach
is similar to the idea of Efros and Shklovskii \cite{11,TN}. 
They
found a soft Coulomb gap in the doped semiconductor. In our
case a slower dependence of the interaction on the distance
leads to a gap of the finite width. Within the toy-model
the gap disappears at the even integer $\kappa$. Although
this result is most probably an artifact of the model, we
expect that the gap is minimal at the even values of
$\kappa$. An interesting question concerns the role of the
collective excitations. For the problem of Coulomb blockade
it was recently discussed in Ref. \cite{12}.
  
The superconducting film includes regions of the size $\sim\lambda_{eff}$
with correlated positive or negative values of the random potential.
The behavior of the vortices
near the borders of the regions is relevant for the transport
properties. 
In particular,
an important process for the resistivity is the transport
of the free vortices between the regions with the same
sign of the random potential through the points of intersection
of the borders. 
Another important process is the transport along the borders,
since the borders constitute a percolating equipotential cluster.
The resistivity depends on the temperature, potential barriers
and concentration of the unbound vortices. It increases
as the number of the unbound vortices grows. At low temperatures
a complicated energy landscape may lead to the glassy dynamical 
behavior.

In conclusion, we obtain that both, the concentrations of
the bound and unbound vortices, increase as the magnetization
of a dot increases. The concentration of the bound vortices depends
on the magnetization in a step-like way and is proportional to the
density of the dots. The concentration of the unbound vortices
is proportional to $1/(\lambda_{eff}a)$. Its dependence on the
magnetic moment of a dot can be represented as oscillations
about a value, proportional to the dot magnetization. 
The period of the oscillations is the same as the width of a step
of the bound vortices concentration. 

This work was partly supported by grants DEFG03-96ER45598, 
NSF DMR-97-05182, THECB ARP 010366-003.
The research of DEF was partly  supported by the Russian Program of Leading
Scientific Schools, grant 96-15-96756.


\end{document}